\documentclass[a4paper,11pt]{article}
\usepackage[margin=0.7in]{geometry}
\usepackage{amsmath}
\usepackage{amsfonts}
\usepackage{mathtools}
\usepackage{amssymb}
\usepackage{graphicx,caption}
\usepackage{epstopdf}
\usepackage{natbib}
\usepackage[explicit]{titlesec}
\usepackage{xcolor}

\bibpunct{(}{)}{;}{a}{}{,}
\pdfoutput=1
\usepackage{multirow}

\begin{document} 

\begin{center}
\LARGE Driving white dwarf metal pollution through unstable eccentric periodic orbits\\
\end{center}

 \begin{center}
	\Large Kyriaki I. Antoniadou$^1$ and Dimitri Veras$^{2,3*}$
 \end{center}
	\begin{flushleft}
	$^1$NaXys, Department of Mathematics, University of Namur, 8 Rempart de la Vierge, 5000 Namur, Belgium\\	
					$^2$Department of Physics, University of Warwick, Coventry CV4 7AL, UK \\					
						$^3$Centre for Exoplanets and Habitability, University of Warwick, Coventry CV4 7AL, UK \\					
						$^*$STFC Ernest Rutherford Fellow \\
\end{flushleft}

\begin{center}
\large{Email: kyriaki.antoniadou@unamur.be}
\end{center}

\date{}

\begin{center}
The final publication is available at\\ https://doi.org/10.1051/0004-6361/201935996
\end{center}

\begin{abstract}
Planetary debris is observed in the atmospheres of over 1,000 white dwarfs, and two white dwarfs are now observed to contain orbiting minor planets. Exoasteroids and planetary core fragments achieve orbits close to the white dwarf through scattering with major planets. However, the architectures that allow for this scattering to take place are time-consuming to explore with $N$-body simulations lasting $\sim 10^{10}$ yr; these long-running simulations restrict the amount of phase space that can be investigated.
Here we use planar and three-dimensional (spatial) elliptic periodic orbits, as well as chaotic indicators through dynamical stability maps, as quick scale-free analytic alternatives to $N$-body simulations in order to locate and predict {\it \textup{instability}} in white dwarf planetary systems that consist of one major and one minor planet on very long timescales. We then classify the instability according to ejection versus collisional events. 
We generalized our previous work by allowing eccentricity and inclination of the periodic orbits to increase, thereby adding more realism but also significantly more degrees of freedom to our architectures. We also carried out a suite of computationally expensive 10 Gyr $N$-body simulations to provide comparisons with chaotic indicators in a limited region of phase space. 
We compute dynamical stability maps that are specific to white dwarf planetary systems and that can be used as tools in future studies to quickly estimate pollution prospects and timescales for one-planet architectures. We find that these maps also agree well with the outcomes of our $N$-body simulations. 
As observations of metal-polluted white dwarfs mount exponentially, particularly in the era of {\it Gaia}, tools such as periodic orbits can help infer dynamical histories for ensembles of systems.
\end{abstract}

{\bf keywords} celestial mechanics -- minor planets, asteroids: general -- Kuiper belt: general -- stars: white dwarfs -- planets and satellites: dynamical evolution and stability -- chaos

%

\section{Introduction}

Observations of white dwarf (WD) exoplanetary systems are abundant and are fast outpacing theoretical efforts to keep up \citep{veras2016b}. Planetary remnants are now observed in the photospheres of between one-quarter and one-half of Milky Way white dwarfs \citep{zucetal2003,zucetal2010,koeetal2014}, over 40 planetary debris disks have been observed to orbit white dwarfs \citep{zucbec1987,ganetal2006,farihi2016}, one exoasteroid is in the process of disintegrating around WD 1145+017 \citep{vanetal2015}, and a dense planetary core fragment has been found to orbit SDSS J1228+1040 \citep{manetal2019}. Furthermore, the number of polluted white dwarfs may increase by an order of magnitude over the coming years; this jump will accompany the corresponding increase in known white dwarfs from {\it Gaia} in 2018 alone \citep{genetal2019}.

The question then is where the major planets are. Almost certainly, they are hidden from view because white dwarfs are so dim. Nevertheless, the final {\it Gaia} data release should reveal at least a dozen planets orbiting white dwarfs through astrometry \citep{peretal2014}. Moreover, Transiting Exoplanet Survey Satellite (TESS), Large Synoptic Survey Telescope (LSST), and even Laser Interferometer Space Antenna (LISA) have the capability of finding planets that orbit white dwarfs \citep{corkip2018,lunetal2018,steetal2018,tamdan2018}.

The location of these planets is the next question. In our solar system, at least Mars, Jupiter, Saturn, Uranus and Neptune will all survive the Sun's transition through the giant branch phases into a white dwarf \citep{schsmi2008,veras2016a}. Their semimajor axes will roughly double as a result of mass loss \citep{omarov1962,hadjidemetriou1963,veretal2011,veretal2013a,doskal2016a,doskal2016b}. They will accompany a sea of debris produced from supercharged Yarkovsky – O'Keefe – Radzievskii – Paddack (YORP)-induced breakup in the asteroid belt \citep{veretal2014} as well remaining Kuiper belt objects that were orbitally shifted through enhanced Yarkovsky effects \citep{veretal2015,veretal2019}.

Changes in stability boundaries during the giant branch phase of evolution can trigger gravitational scattering \citep{debsig2002} or delay it for $10^8, 10^9$ , or even $10^{10}$ yr \citep{veretal2013b,veretal2016,veretal2018,musetal2014,musetal2018,vergae2015}, particularly when secular and mean motion resonance (MMR) sweeping occurs \citep{bonetal2011,debes12,voyetal2013,freha14,smaetal2018,smamar2019}. In at least enough cases for the outcome to be readily observable, the scattering results in small bodies that intersect the white dwarf Roche radius from distances of several au. The subsequent breakup forms a disk \citep{graetal1990,jura2003,debes12,beasok2013,veretal2014,broetal2017,veretal2017} that is eventually accreted onto the white dwarf atmosphere and thereby pollutes it.

The authors of several of the references mentioned above performed simulations that ran only on timescales of gigayears or shorter because of natural computational limitations. However, old polluted white dwarfs are common \citep{holetal2018}, where in this context ``old'' refers to white dwarfs with cooling ages of at least 5-8~Gyr. Furthermore, full-lifetime simulations that can achieve these timescales \citep[e.g.,][]{veretal2016} have been understandably limited to exploring a narrow band of phase space.

In order to address this difficulty, \cite{wd1} presented an analytical alternative to $N$-body simulations by exploring the predictive power of periodic orbits and chaotic indicators through dynamical stability maps (DS maps). They did so in the context of the planar circular restricted three-body problem (CRTBP) with an outer planet (on a circular orbit) and inner asteroid near or inside of the 2:1 MMR. They chose the 2:1 MMR because MMRs have been shown to represent dynamical mechanisms that lead to white dwarf pollution \citep{bonetal2011,voyetal2013,freha14,smaetal2018,smamar2019}, the 2:1 MMR is a strong first-order resonance, and the 2:1 MMR in particular was a focus of \cite{debes12}, allowing for comparisons to be made.

Here, we significantly extend \cite{wd1} by instead considering the elliptic restricted three-body problem (ERTBP) as well as, in a more limited capacity, the spatial circular and elliptic restricted three-body problems (3D-CRTBP and 3D-ERTBP).

The ERTBP is more realistic to consider than the CRTBP in white dwarf planetary systems. Even a massive planet on an exactly circular orbit on the main sequence will attain some eccentricity, however small, as a result of giant branch mass loss \citep{veretal2011}. The ERTBP will also allow us to make a direct comparison to previous single-planet scattering studies, such as that of \cite{debes12}, which adopted a planetary eccentricity equivalent to the current eccentricity of Jupiter (0.048). The eccentricity of the asteroid through scattering can be excited to the required level ($e > 0.99$) by a variety of architectures, as demonstrated in the numerical studies listed above, as well as through specific MMRs like the 4:1 \citep{picetal2017}. Furthermore, a second major planet in the system can excite the other planet's eccentricity to any level \citep{caretal2019}.

In Sect. \ref{model} we provide our model and the basic notions of periodic orbits, linear stability, and the MMR. In Sect. \ref{tools} we show the predictive power of the periodic orbits, which can shape the phase space into different domains of stable or chaotic motion. In Sect. \ref{nbody} we introduce the setup of the $N$-body integrations, and in Sect. \ref{res} we present the results of the $N$-body simulations. Finally, we summarize our findings in Sect. \ref{con}.

\section{Model setup, periodic orbits, and MMRs}\label{model}
\subsection{Restricted three-body problem}
In the framework of the planar restricted three-body problem (2D-RTBP), we considered a system consisting of a star, an inner asteroid, and an outer planet of masses $m_{\rm S}$, $m_{\rm A}=0$ and $m_{\rm P}=m_{\rm J}=0.001 m_{\odot}$, respectively, and the problem parameter $\mu=\frac{m_{\rm P}}{m_{\rm S}+m_{\rm P}}$. The asteroid (massless body) does not affect the motions of the star and the planet (primaries) but rather moves under their gravitational attraction. In the 2D-RTBP, we defined a suitable planar rotating frame of reference, $Oxy$, that is centered at the center of mass of the primaries \citep[see, e.g.,][]{hadj75}.

In the spatial RTBP (3D-RTBP), the asteroid was allowed to evolve on an inclined orbit with regard to the plane of the orbits of the primaries. The spatial rotating frame, $Gxyz$, has an origin that coincides with the center of mass of the massive bodies, its $Gz$-axis is perpendicular to the $Gxy$-plane, to which the motion of the WD and the planet is confined \citep[see, e.g.,][for more details on dynamics of the 3D-RTBP]{spa}.   

Whether in a planar or a spatial rotating frame, the planet was allowed to evolve on a circular $e_{\rm P}=0$ (2D-CRTBP for a planar or 3D-CRTBP for a rotating frame) or an elliptic $e_{\rm P}>0$ (2D-ERTBP for a planar or 3D-ERTBP for a rotating frame) orbit. More information about the theory of the RTBPs can be found in \citet{sze}. 

We studied periodic orbits under the normalization that sets the sum of the masses and the gravitational constant, $G$, equal to unity, that is, $m_{\rm S}+m_{\rm A}+m_{\rm P}=1$ (or $m_{\rm S}=1-m_{\rm P}$ and hence, $\mu$=0.001) and $G=1$. 

When viewed in an inertial frame of reference, the orbits of the asteroid and planet correspond to Keplerian ellipses that are characterized by the osculating elements \citep[see, e.g.,][]{murray}. The notation we use is $a$ (semimajor axis), $e$ (eccentricity), $i$ (inclination), $\omega$ (argument of pericenter), $M$ (mean anomaly), and $\Omega$ (longitude of ascending node), and $\varpi=\omega+\Omega$ (longitude of pericenter), $\Delta\varpi$ (apsidal difference), and $\lambda=\varpi+M$ (mean longitude). 

As we study an interior MMR, the massless body (asteroid) evolves in an inner orbit with regard to that of the planet, namely $a_{\rm A}<a_{\rm P}$. During the computation of the periodic orbits, we always set $a_{\rm P}=1$, without loss of generality, so that the period of the planet is $T_0=2\pi$.

\subsection{Periodic orbits and stability}
When we consider a set of positions and velocities, for instance, $\textbf{Q}(t)$, in the phase space of the primaries, then an orbit is periodic if it fulfills the periodic conditions $\textbf{Q}(0) = \textbf{Q}(T)$, where $T$ is the orbital period with $t = kT$, $k \ge 1 \in \mathbb{Z}$. A periodic orbit is symmetric if it is invariant under the fundamental symmetry $\Sigma: (t,x,y) \rightarrow (-t,x,-y)$ \citep{hen}. If the periodic orbit is not invariant, it is asymmetric and $\Sigma$ will map it to its mirror image. When we take initial conditions on a Poincar\'e surface of section \citep{pnc} of the suitable rotating frame, $Oxy$ or $Gxyz$, the periodic orbits are the fixed points of the Poincar\'e map. They can then form a family through different schemes of monoparametric continuation. The origin and continuation of periodic orbits along with the periodicity conditions they should fulfill are described by \citet{spis} for the 2D-RTBPs and by \citet{spa} for the 3D-RTBPs.

The periodic orbits shape the phase space through their linear stability. We used the conjugate eigenvalues (in reciprocal pairs) of the monodromy matrix of the variational equations of the system in order to classify them as stable or unstable. In general, a periodic orbit is considered stable if and only if all of the eigenvalues lie on the unit circle \citep[see, e.g.,][for different types of instabilities]{broucke69,marchal90}. At the orbits where a transition of stability takes place, we have bifurcation points that generate asymmetric periodic orbits. We here plot the stable periodic orbits in blue and the unstable orbits in red. The motion in the domains around the stable orbits is regular, and the long-term stability can be guaranteed thereby. However, in the vicinity of the unstable orbits, chaos emerges and instability events occur, such as collisions or escape. 

An intrinsic property of the planar periodic orbits is their vertical stability \citep{hen}. The vertically stable planar periodic orbits can guarantee the stability of systems with small mutual inclinations around them. The transition of vertical stability to instability along a family is represented by the vertical critical orbit (v.c.o.), which is a bifurcation point that generates spatial periodic orbits. The vertical stability along the planar families is here depicted by solid lines and the vertical instability by dashed lines.

In order to determine the extent of the stable (and/or unstable) regions in the phase space of complex systems, we computed DS maps, which provide a visual representation of the domains. In this respect, we used chaotic indicators, which can help distinguish chaos from order for each of the initial conditions of a DS map. The detrended fast Lyapunov indicator (DFLI) \citep[see][for the definition of the FLI]{froe97}, which is the FLI divided by time \citep{voy08}, is fast and accurate in tracing chaoticity.  When the orbit is regular, the DFLI remains almost constant over time (with values lower than 10). However, when the orbit is chaotic, then the DFLI increases exponentially. We computed the DFLI for a maximum integration time, $t_{max}=250$ kyr, which has been proved to be adequate for revealing chaos in the RTBP with the masses we use. In the DS maps, dark colors depict regular orbits, and pale colors highlight traced instability. White shows the failure of the numerical integration at $t<t_{max}$, which is caused by a small integration step during very close encounters between the bodies. The numerical integration stops when the threshold $10^{30}$ is reached by the DFLI.

\subsection{MMR}
In an MMR two proper frequencies of the system become commensurable. An MMR associates the ratio of the mean motions, $n_i$, $i$ standing either for the asteroid, ${\rm A}$, or the planet, ${\rm P}$, with the rational ratio of two integers, $q,p\neq 0 \in \mathbb{Z}$, namely $\frac{n_{\rm P}}{n_{\rm A}}=(\frac{a_{\rm A}}{a_{\rm P}})^{-\frac{3}{2}}\approx\frac{p+q}{p}$, where q is the order of the resonance. A periodic orbit in the rotating frame shows the exact location of an MMR in phase space.

We here discuss the 2:1 interior MMR, and because the periodic orbits correspond to the stationary solutions of an averaged Hamiltonian, we can define the resonant angles on which they depend as follows:
\begin{equation}\begin{array}{l}
\theta_1=\lambda_{\rm A}-2\lambda_{\rm P}+\varpi_{\rm A} \;\; {\rm and} \;\;
\theta_2=\lambda_{\rm A}-2\lambda_{\rm P}+\varpi_{\rm P} \\
\end{array}
.\end{equation}

In the neighborhood of a stable periodic orbit in phase space, the resonant angles and the apsidal difference librate. When all of them librate about 0 or $\pi$, the evolution takes place in the vicinity of a symmetric periodic orbit. On the other hand, when they librate about other angles, the periodic orbit is asymmetric. In the neighborhood of unstable periodic orbits, these angles rotate. Given the values of the mean anomalies (0 or $\pi$) and the different combinations we can obtain with aligned ($\Delta\varpi=0$) or anti-aligned orbits ($\Delta\varpi=\pi$), we finally defined four different symmetric configurations: ($\theta_1,\theta_2)=(0,0), (0,\pi), (\pi,0),$ and $(\pi,\pi)$. In order to distinguish them on the eccentricities plane, we use the following sign-convention. We assign a positive (or negative) value to $e_{\rm A}$ when $\theta_1$ equals 0 (or $\pi$) in the family of each configuration. The same accordingly holds for $e_{\rm P}$ and $\theta_2$.

Nonetheless, there also exist other mechanisms that can protect the systems from disruption. For instance, we can observe the secondary resonance inside an MMR, where we consider the average of the frequencies between rotation and libration of $\theta_2$ and $\theta_1$, respectively \citep{moons93}. There also exists the apsidal resonance for the nonresonant orbits, where only the apsidal difference oscillates about 0 or $\pi$ and the rest of the resonant angles rotate \citep{murray,morbidelli2002}. For the resonant orbits we can have apsidal resonance only for eccentric orbits that are asymmetric \citep{mich08}. Moreover, stability can be attained by apsidal difference oscillation during the transition between two symmetric configurations inside an MMR, where the passage is merely kinematical \citep{mich08}. A systematic study of the regions around the periodic orbits in relation to the above mechanisms of protection of the phases can be found in \citet{kiaasl} for the 2:1 MMR and \citet{spis} for the 3:2, 5:2, 3:1, 4:1, and 5:1 MMRs.

\section{Periodic orbits and DS maps as diagnostic tools}\label{tools}
\subsection{2D-RTBPs}\label{2dmaps}
The planar families of periodic orbits in 2:1 MMR when $e_{\rm P}>0$ (2D-ERTBP) for every configuration were computed by \citet{kiaasl}. We here first employed them to select the symmetric periodic orbits, which then acted as diagnostic tools for the computation of the four DS maps (one for each configuration) on the $(e_{\rm A},e_{\rm P})$ plane. 

\begin{figure}
\centering
\resizebox{.425\hsize}{!}{\includegraphics{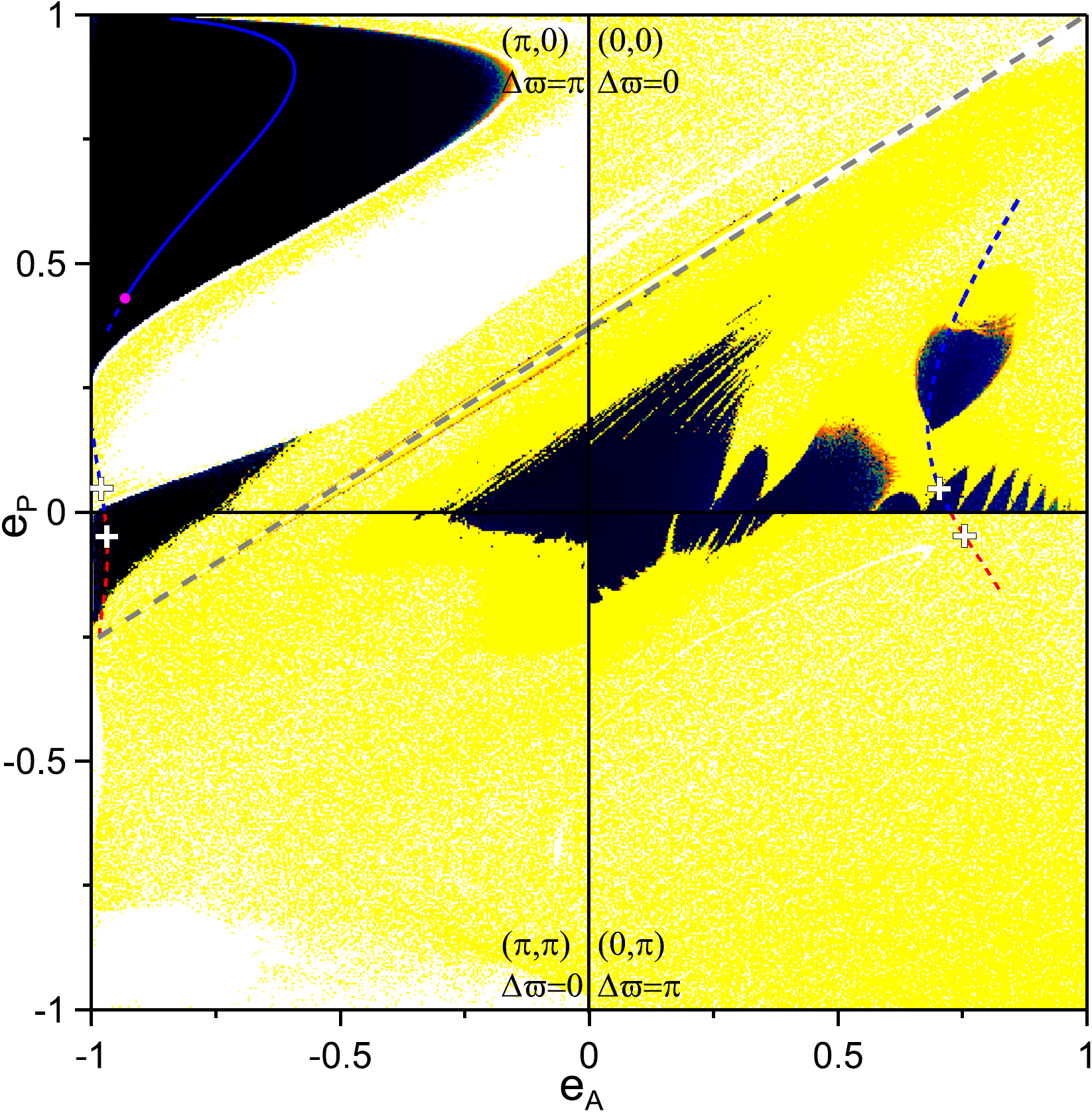}}\\
\includegraphics[width=3.5cm,height=0.7cm]{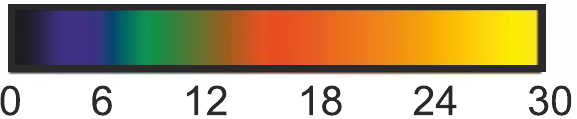} 
\caption{Families of symmetric periodic orbits in 2:1 MMR of the 2D-ERTBP when $m_{\rm P}=m_{\rm J}$ are overplotted on the DS maps of the plane $(e_{\rm A},e_{\rm P})$. The white crosses represent the chosen periodic orbits for our study,  the orbital elements of which remain fixed for the computation of each map. We use a positive (or negative) value for $e_{\rm A}$ when $\theta_1$ equals 0 (or $\pi$) in the family of each configuration. The same convention applies to $e_{\rm P}$ and $\theta_2$. Dark (pale) regions correspond to stable (chaotic) domains in phase space as attributed by the logarithmic values of the DFLI shown by the colored bar. Blue (red) curves stand for the horizontal stability (instability) of the periodic orbits, while the solid (dashed) curves show the vertical stability (instability). The magenta dot corresponds to a transition of the vertical stability. The dashed gray curve depicts the collision line between the planet and the asteroid.}
\label{21e_maps}
\end{figure}

Particularly, as we first aim to explore WD pollution when $e_{\rm P}=e_{\rm J}=0.048$, the chosen symmetric periodic orbits correspond to the following initial conditions per symmetric configuration: 
\begin{itemize}
        \item ($\theta_1,\theta_2)=(0,0)$ and $\Delta\varpi=0$:\\ $e_{\rm A}=0.7035$, $a_{\rm A}=0.6127$, $\varpi_{\rm A}=M_{\rm A}=\varpi_{\rm P}=M_{\rm P}=0{^{\circ}}$
        \item ($\theta_1,\theta_2)=(0,\pi)$ and $\Delta\varpi=\pi$:\\ $e_{\rm A}=0.7542$, $a_{\rm A}=0.6003$, $\varpi_{\rm A}=M_{\rm P}=180{^{\circ}}$ and $\varpi_{\rm P}=M_{\rm A}=0{^{\circ}}$
        \item ($\theta_1,\theta_2)=(\pi,0)$ and $\Delta\varpi=\pi$:\\ $e_{\rm A}=0.9810$, $a_{\rm A}=0.6302$, $\varpi_{\rm A}=M_{\rm A}=180{^{\circ}}$ and $\varpi_{\rm P}=M_{\rm P}=0{^{\circ}}$
        \item ($\theta_1,\theta_2)=(\pi,\pi)$ and $\Delta\varpi=0$:\\ $e_{\rm A}=0.9694$, $a_{\rm A}=0.6307$, $\varpi_{\rm A}=\varpi_{\rm P}=0{^{\circ}}$ and $M_{\rm A}=M_{\rm P}=180{^{\circ}}$  
\end{itemize}

\begin{figure}
\centering
\resizebox{0.4\hsize}{!}{\includegraphics{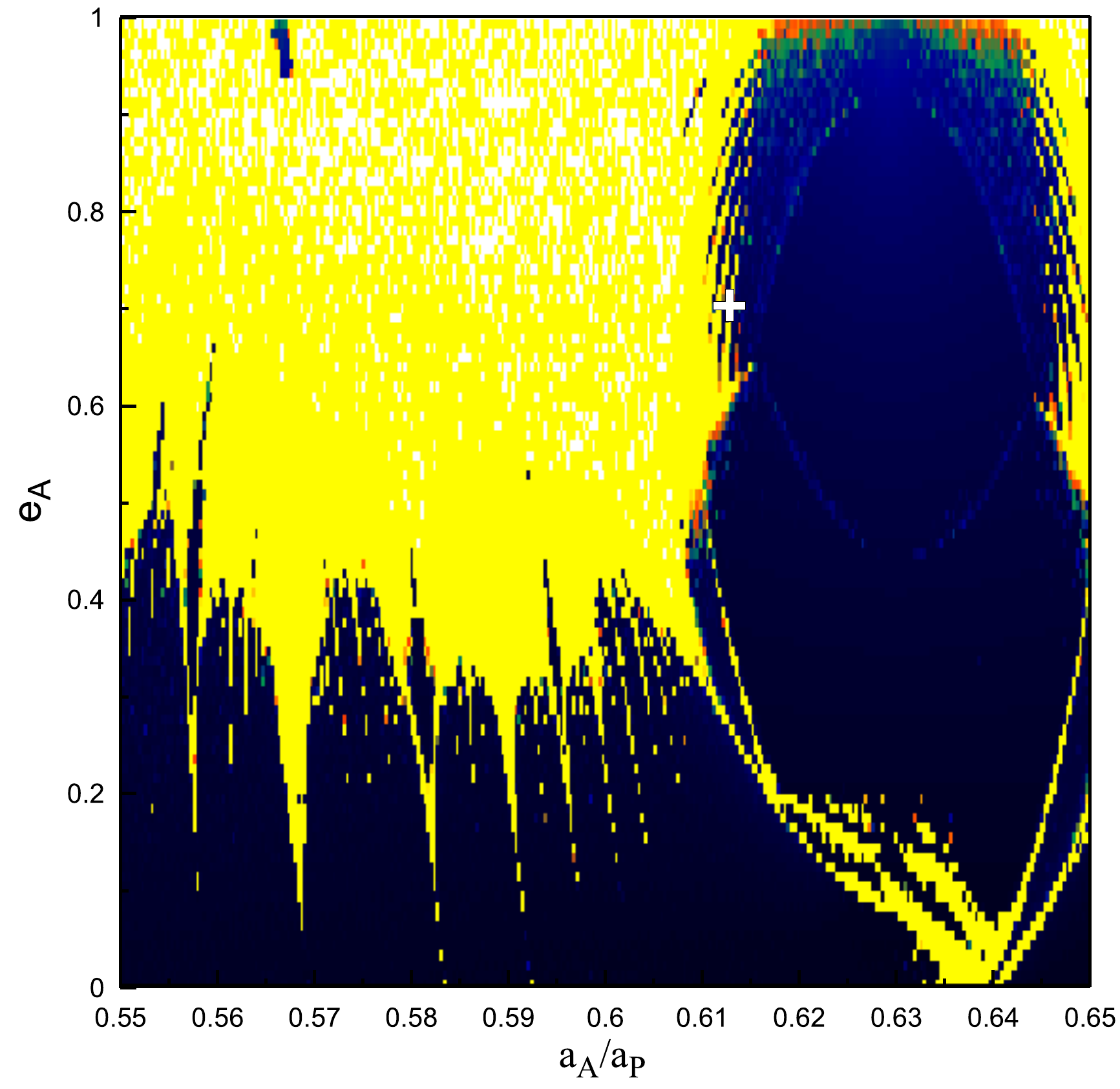}}\\
\resizebox{0.4\hsize}{!}{\includegraphics{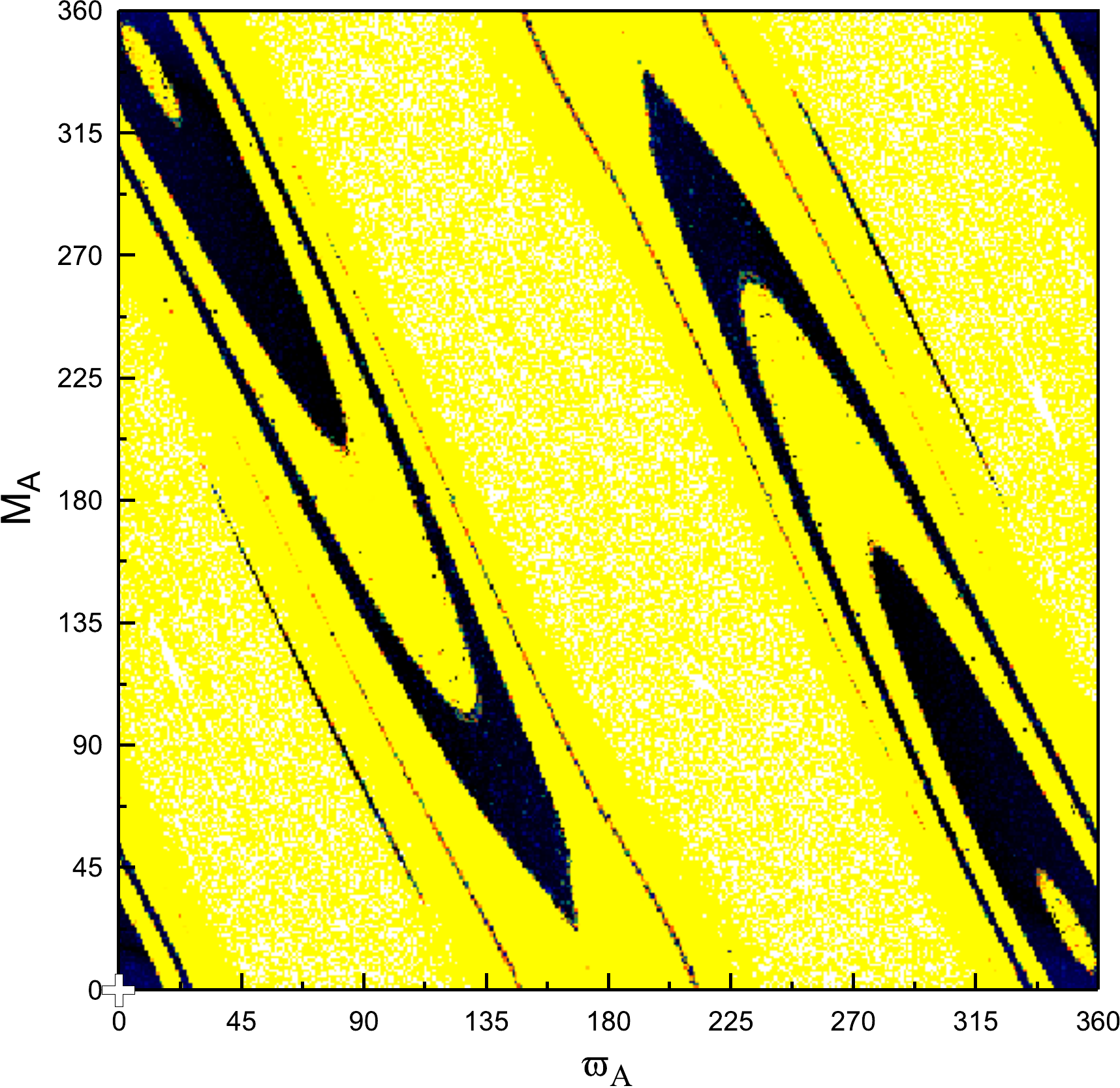}}\\
\resizebox{0.4\hsize}{!}{\includegraphics{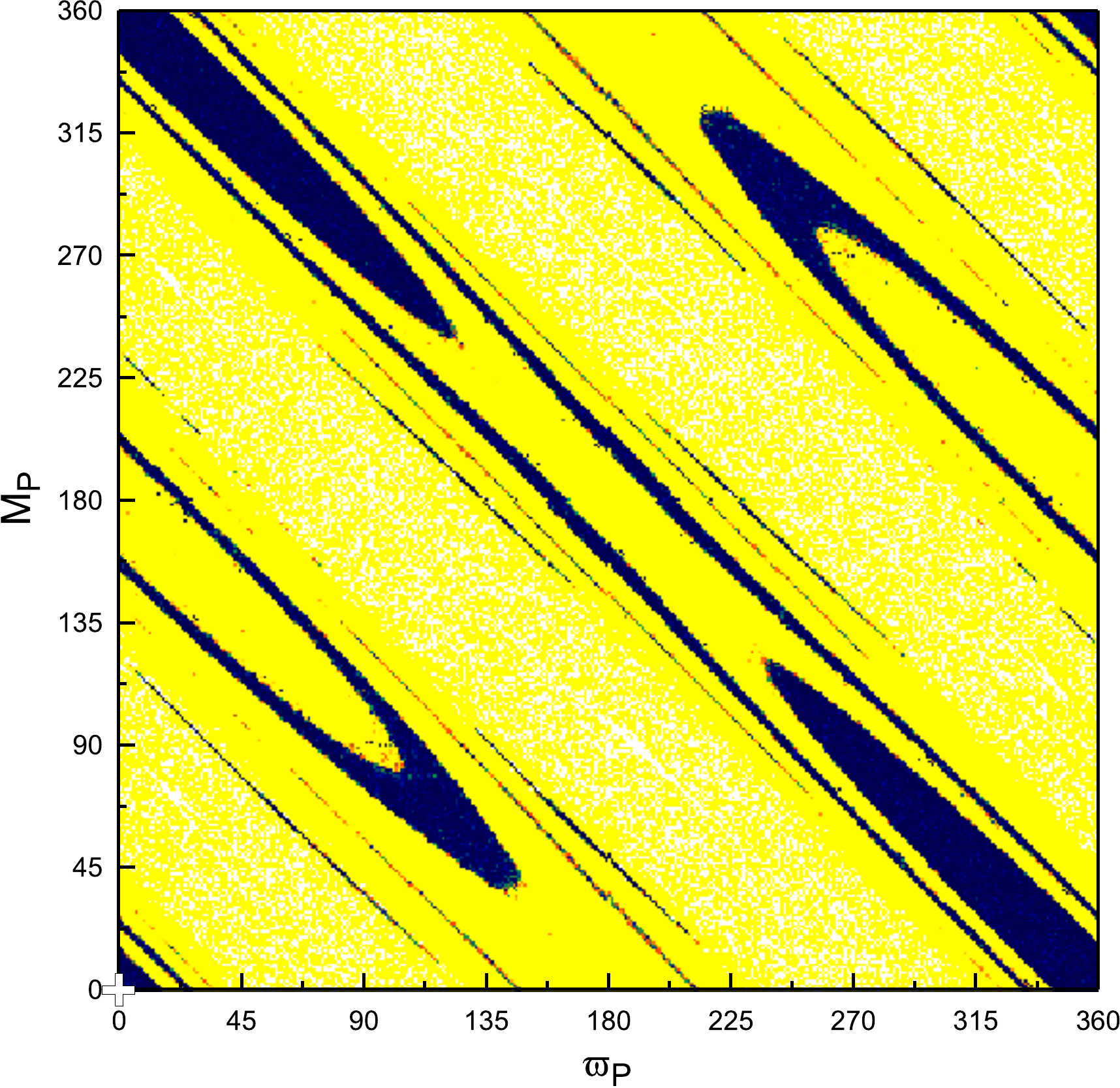}}\vspace{-0.2cm}
\caption{DS maps on the planes $(a_{\rm A}/a_{\rm P},e_{\rm A})$ (top), $(\varpi_{\rm A},M_{\rm A})$ (middle), and $(\varpi_{\rm P},M_{\rm P})$ (bottom) for the symmetric configuration ($\theta_1,\theta_2)$=$(0,0)$ and $\Delta\varpi$=0 presented as in Fig. \ref{21e_maps}.}
\label{conf00}
\end{figure}

\begin{figure}
\centering
\resizebox{0.4\hsize}{!}{\includegraphics{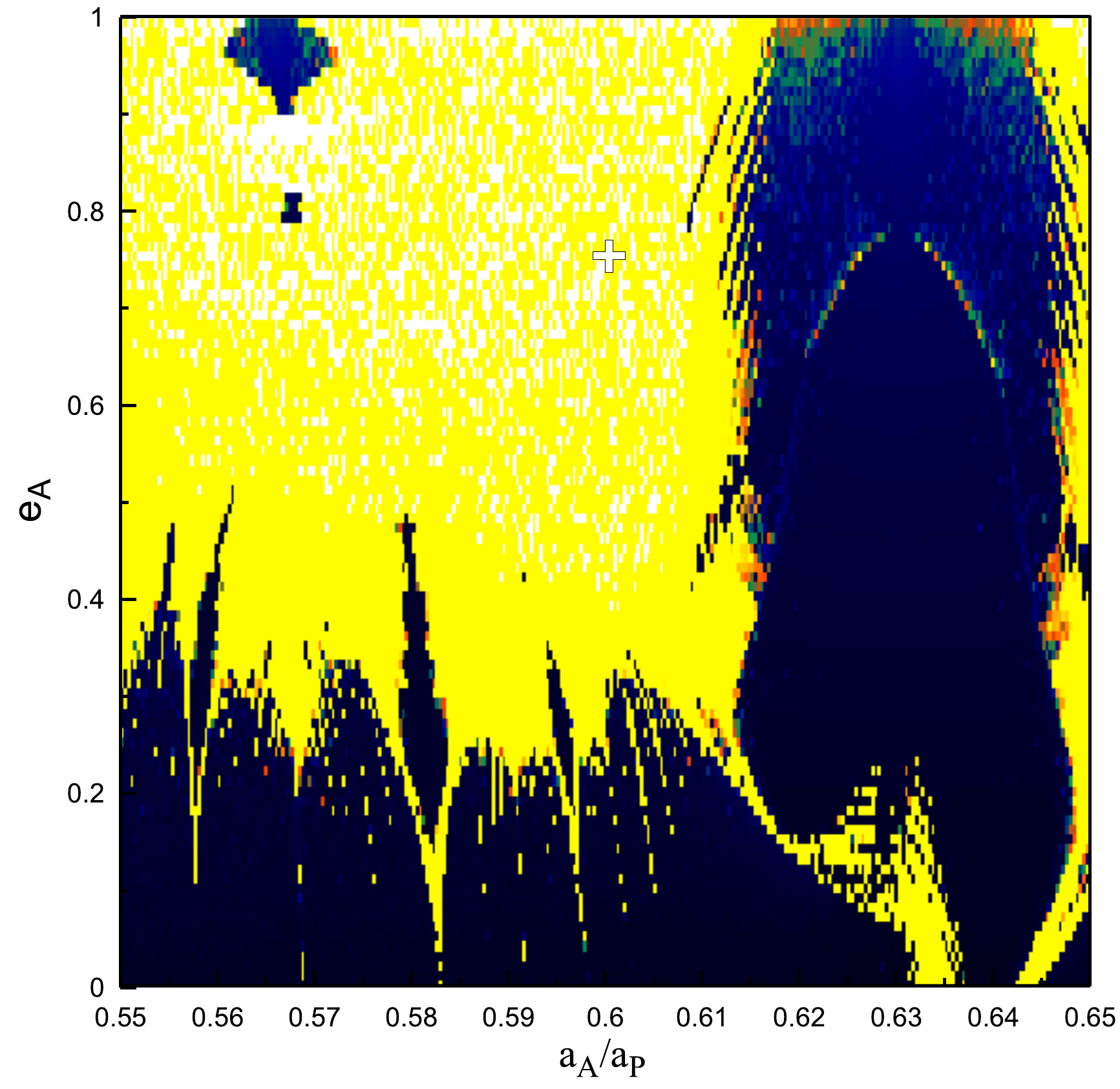}}\\
\caption{DS map on the plane $(a_{\rm A}/a_{\rm P},e_{\rm A})$ for the symmetric configuration ($\theta_1,\theta_2)=(0,\pi)$ and $\Delta\varpi=\pi$.}
\label{conf0p}
\end{figure}

\begin{figure}
\centering
\resizebox{0.4\hsize}{!}{\includegraphics{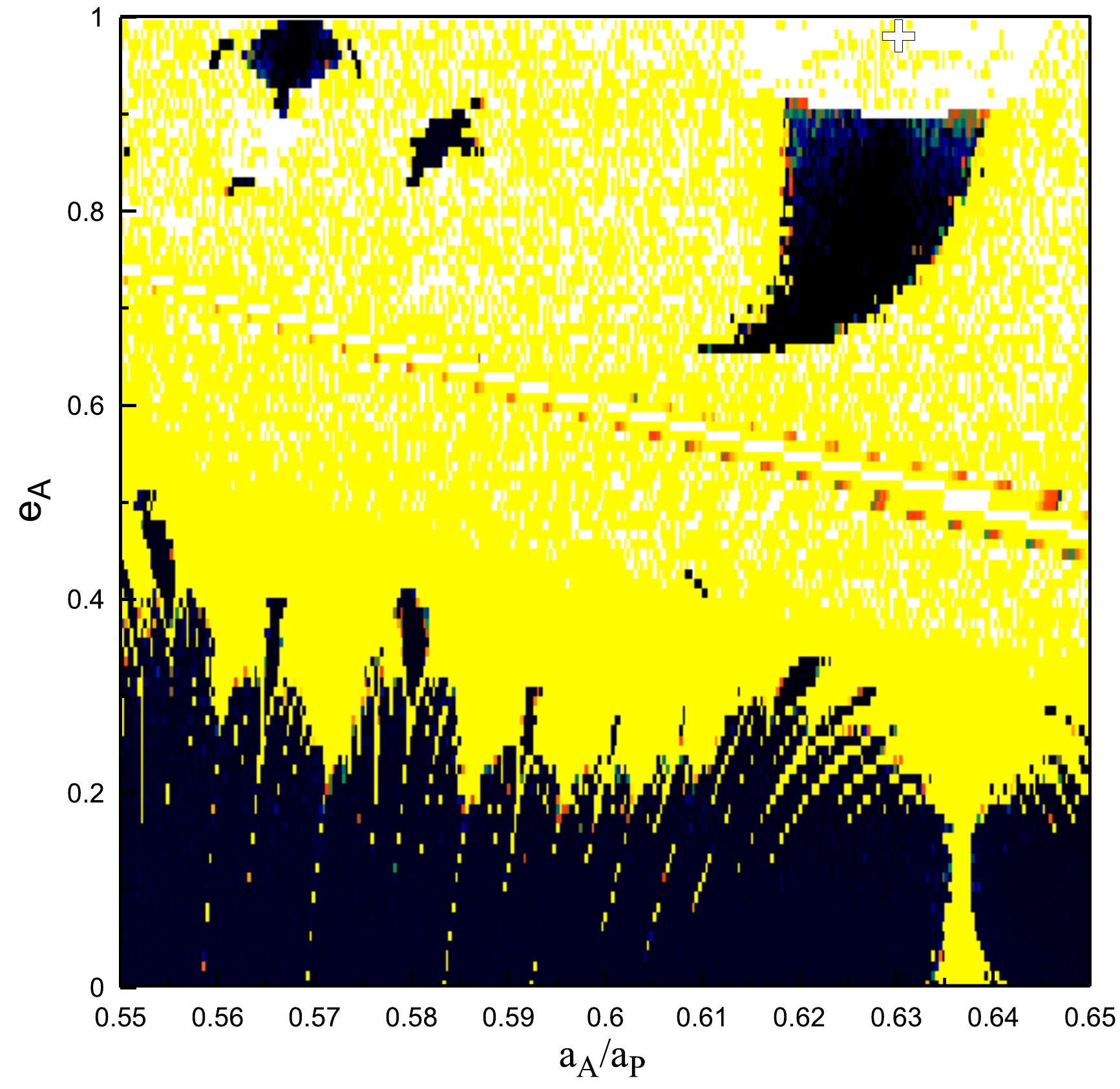}}\\
\resizebox{0.4\hsize}{!}{\includegraphics{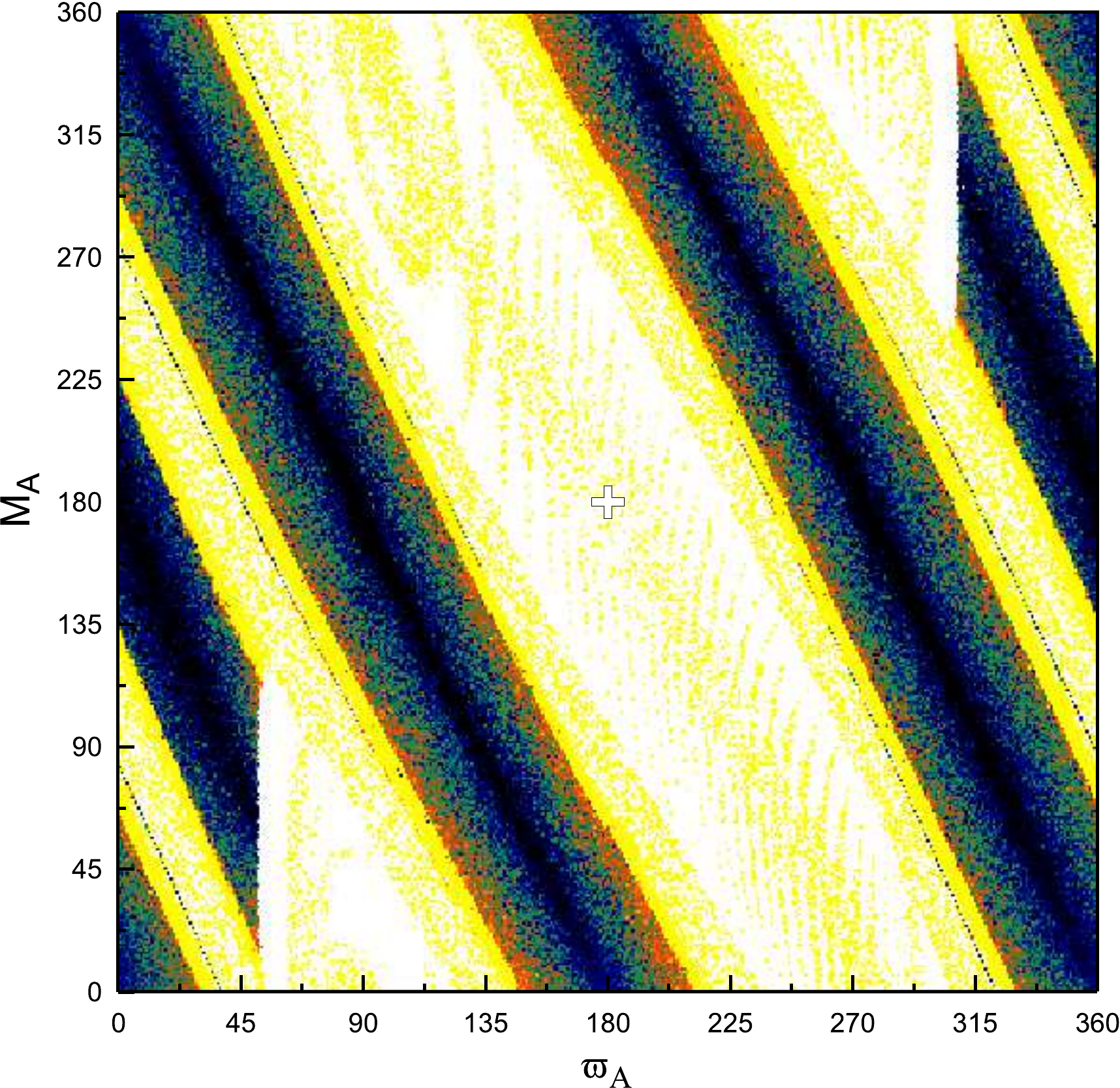}}\\
\resizebox{0.4\hsize}{!}{\includegraphics{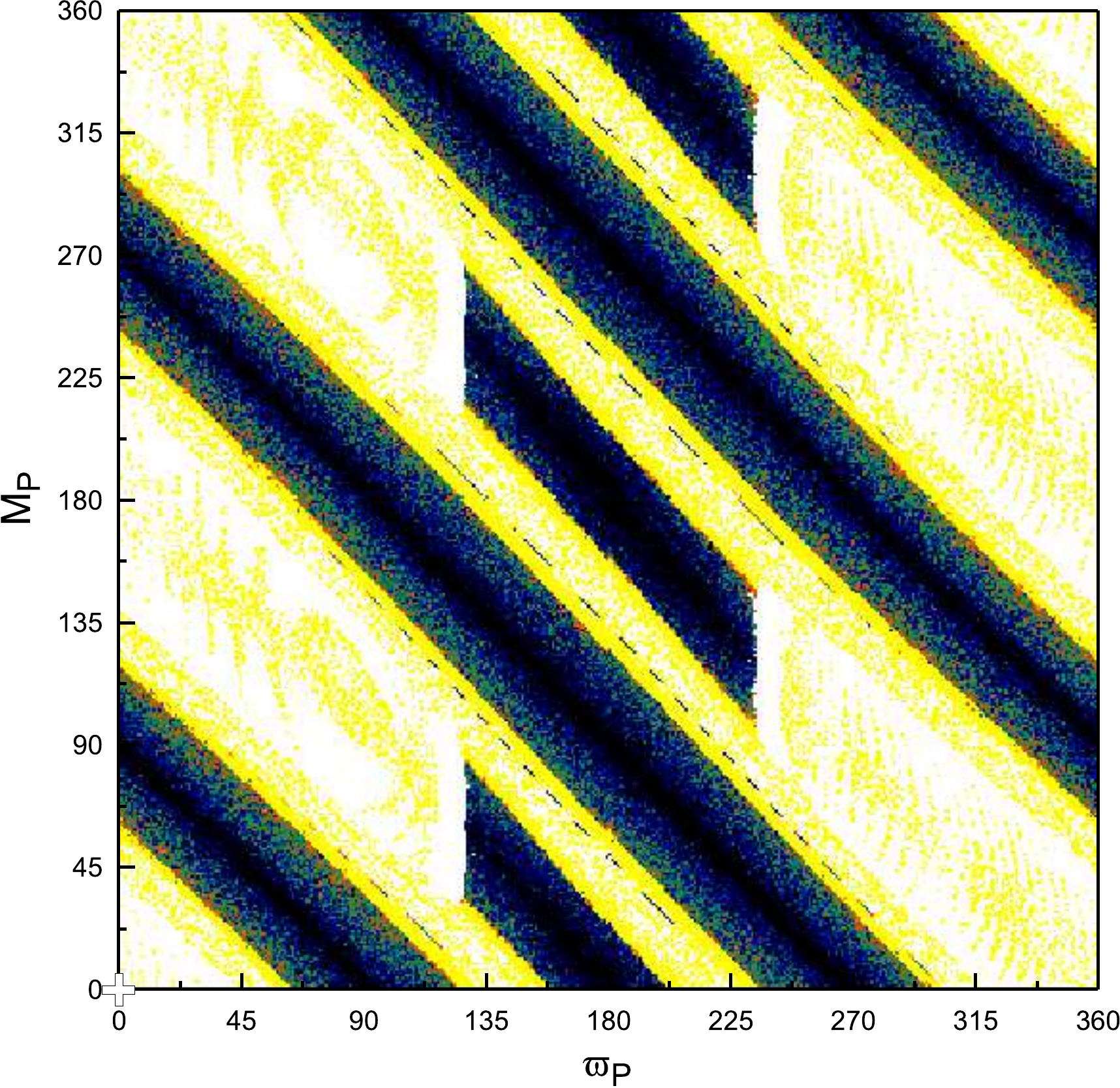}}\vspace{-0.2cm}
\caption{DS maps for the symmetric configuration ($\theta_1,\theta_2)=(\pi,0)$ and $\Delta\varpi=\pi$ presented as in Fig. \ref{conf00}.}
\label{confp0}
\end{figure}

\begin{figure}
\centering
\resizebox{0.4\hsize}{!}{\includegraphics{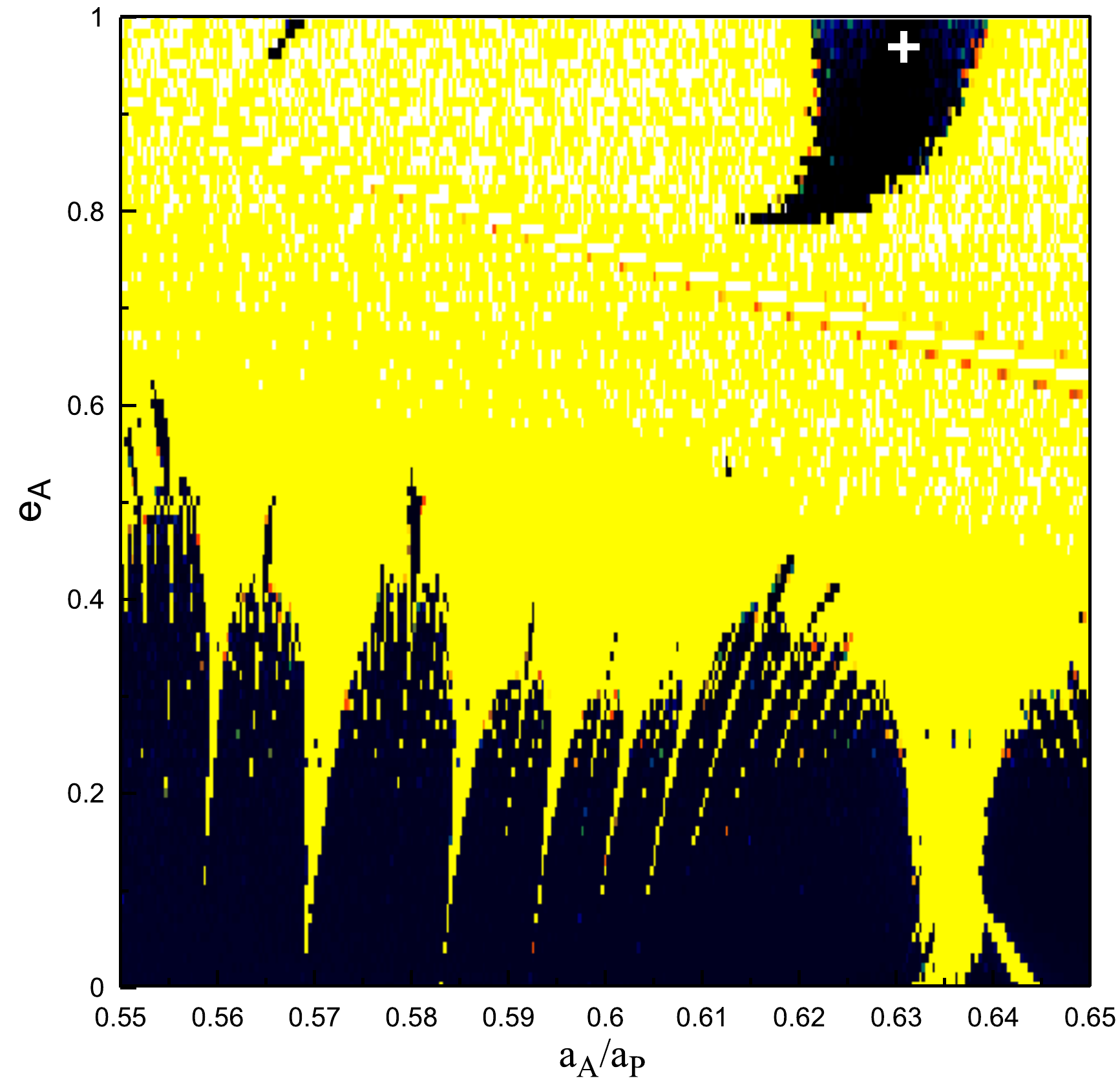}}\\
\resizebox{0.4\hsize}{!}{\includegraphics{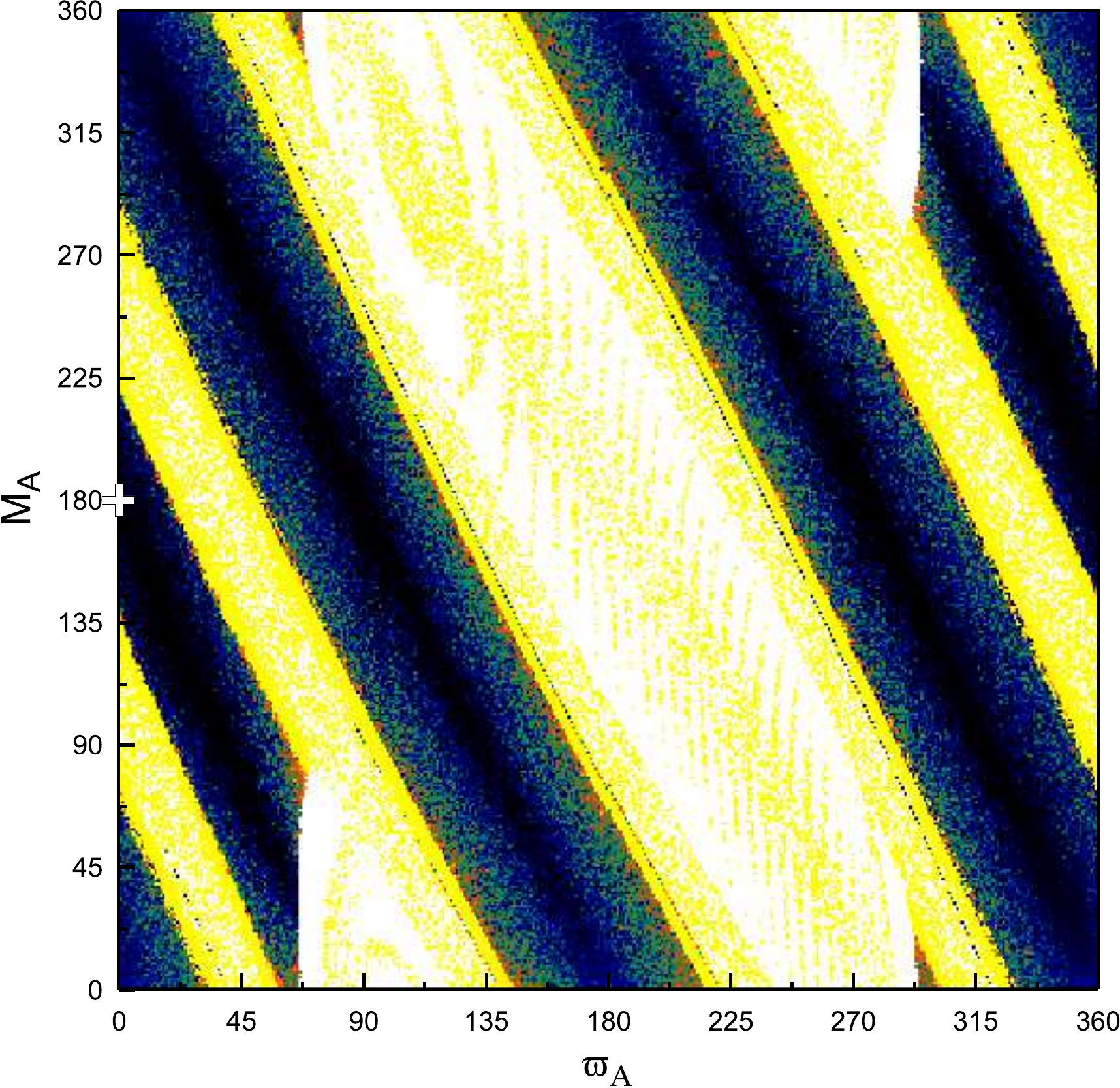}}\\
\resizebox{0.4\hsize}{!}{\includegraphics{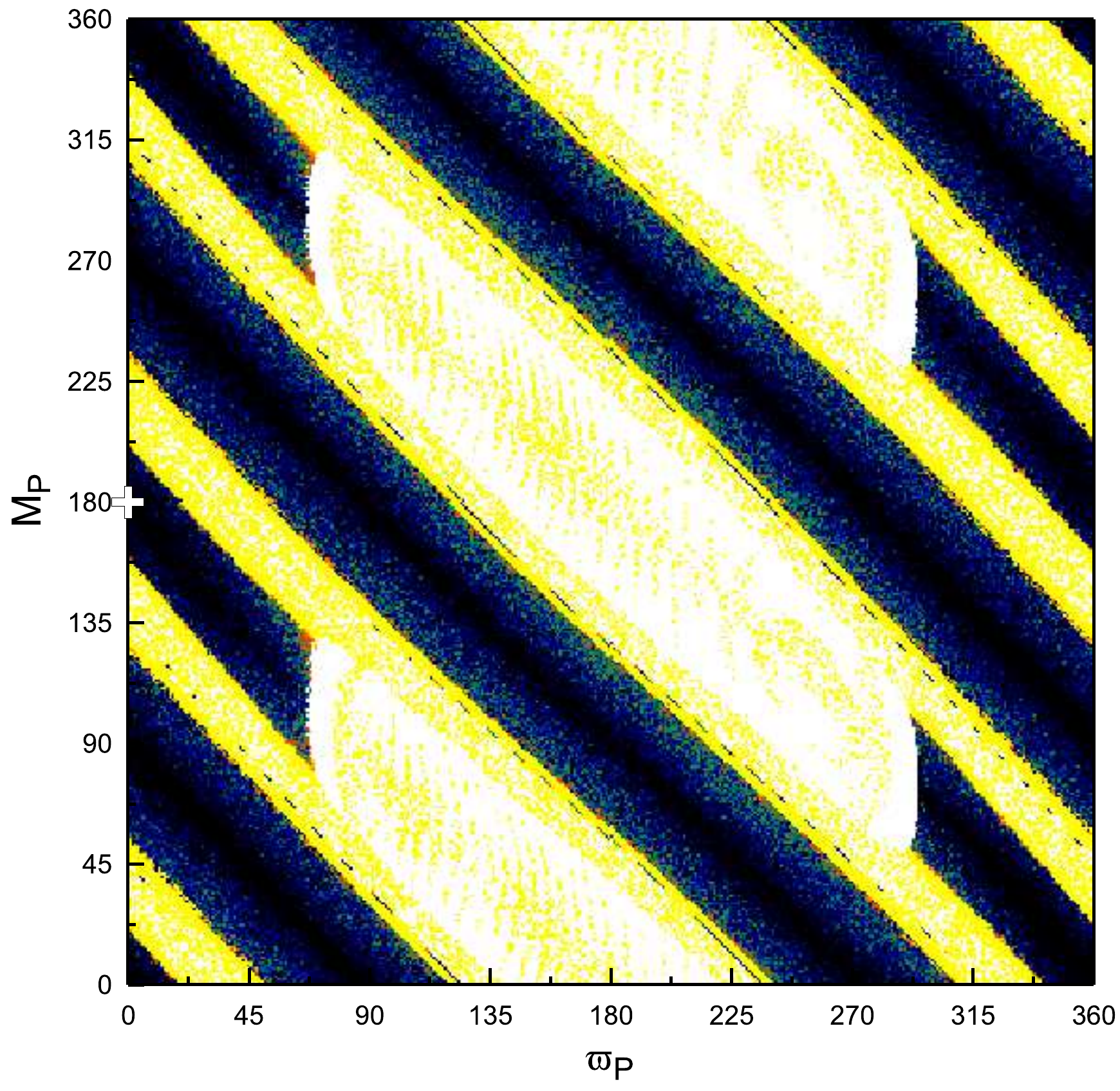}} \vspace{-0.2cm}
\caption{DS maps for the symmetric configuration ($\theta_1,\theta_2)=(\pi,\pi)$ and $\Delta\varpi=0$ presented as in Fig. \ref{conf00}.}
\label{confpp}
\end{figure}

In Fig. \ref{21e_maps} we overplot the families of 2:1 resonant symmetric periodic orbits on the DS maps computed on the $(e_{\rm A},e_{\rm P})$ plane and highlight the chosen orbits with white crosses. When an island of stability is apparent around the selected periodic orbits (for $e_{\rm P}<0.2$), we have a 1:1 secondary resonance where $\theta_1$ librates around 0 or $\pi$ according to the symmetric configuration to which the island belongs. More specifically, in the island below the periodic orbit in configuration $(0,0),$ $\theta_1$ librates about 0, whereas in the configuration ($\pi,\pi$), it librates about $\pi$.

As the degrees of freedom of the architectures studied here are increased compared to those described in \cite{wd1}, the domains in phase space that have to be explored significantly increase as well. Driven by these four periodic orbits, we also computed DS maps on the planes $(a_{\rm A}/a_{\rm P},e_{\rm A})$, $(\varpi_{\rm A},M_{\rm A}),$ and $(\varpi_{\rm P},M_{\rm P})$ and show how the domains around the system studied change for each symmetric configuration (see Figs. \ref{conf00}-\ref{confpp}). 

In particular, we present in Fig. \ref{conf00} the DS maps for the symmetric configuration ($\theta_1,\theta_2)$=$(0,0)$ and $\Delta\varpi=0$. In the small stable domain arising around the periodic orbit, we have 1:1 secondary resonance inside the 2:1 MMR, where $\theta_1$ librates about 0, and long-term stability is expected therein. 

In Fig. \ref{conf0p} we present one DS map for the symmetric configuration ($\theta_1,\theta_2)$=$(0,\pi)$ and $\Delta\varpi=\pi$, that is, only on the plane $(a_{\rm A}/a_{\rm P},e_{\rm A})$, because the DS maps on the planes $(\varpi_{\rm A},M_{\rm A})$ and $(\varpi_{\rm P},M_{\rm P})$ resulted solely in chaotic orbits. The periodic orbit is surrounded by chaotic domains, and instability events are expected to take place in this dynamical neighborhood. 

In Fig. \ref{confp0} we reveal the symmetric configuration $(\theta_1,\theta_2)$=$(\pi,0)$ and $\Delta\varpi=\pi$. Below the periodic orbit (for $e_{\rm A}\approx 0.8$) there exists an island of stability, where the regularity of the orbits is maintained as a result of the 1:1 secondary resonance inside the 2:1 MMR, where the resonant angle $\theta_1$ librates about $\pi$. 

In Fig. \ref{confpp}, where the phase space of the configuration ($\theta_1,\theta_2)$=$(\pi,\pi)$ and $\Delta\varpi$=0 is shown, we have an island of stability around the periodic orbit (for $e_{\rm A}>0.8$), where the 1:1 secondary resonance takes place with $\theta_1$ librating about $\pi$. This is expected to preserve stability for long-time spans.

Notably, on the plane $(a_{\rm A}/a_{\rm P},e_{\rm A})$ of the Figs. \ref{conf00}-\ref{confpp}, an instability region is observed for very low values of $e_A$ at 2:1 MMR. This region is justified by the Poincar\'e surface of sections, which have long proved that the chaotic regions therein are bounded and therefore, despite the irregular motion, escape cannot occur. These regions are also sensitive to the eccentricity of Jupiter, which in the DS maps here is equal to 0.048 (ERTBP), a value that is still very close to zero (see the DS maps in the CRTBP studied by \citealt*{wd1}). We refer to \citet{MoMo93}, \citet{mifm95}, \citet{NeFe97} and \citet{havo01}, for instance, for more details on the dynamics of this particular region.

An additional stable region observed on the plane $(a_{\rm A}/a_{\rm P},e_{\rm A})$ for $a_{\rm A}/a_{\rm P}\approx 0.57$ and $e_A>0.9$ in Figs. \ref{conf0p} and \ref{confp0} is also worth mentioning. This domain is associated with the 7:3 MMR, given the libration of the resonant angles and the apsidal difference. Although this is intriguing, it is beyond the scope of this study.

To our knowledge, asymmetric configurations ($\Delta\varpi\neq 0$ or $\pi$) linked with WD pollution do not exist in the literature. In order to address the issue of geometric asymmetry ($\Delta\varpi=\varpi_{\rm P}-\varpi_{\rm A}$) and dynamical asymmetry ($\Delta M=M_{\rm P}-M_{\rm A}$) of the orbits in our $N$-body simulations and provide a first insight of how these angles can also cause instability events, we use as initial conditions the following asymmetric periodic orbits, which belong to the \textit{Family 1} of \citet{kiaasl}: 
\begin{itemize}
        \item 1:\\ $e_{\rm A}=0.0860$, $e_{\rm P}=0.5580$, $a_{\rm A}=0.6298$, $\varpi_{\rm A}=M_{\rm A}=0{^{\circ}}$, $\varpi_{\rm P}=91.50{^{\circ}}$, $M_{\rm P}=218.65{^{\circ}}$
        \item 2:\\ $e_{\rm A}=0.2000$, $e_{\rm P}=0.6424$, $a_{\rm A}=0.6294$, $\varpi_{\rm A}=M_{\rm A}=0{^{\circ}}$, $\varpi_{\rm P}=322.13{^{\circ}}$, $M_{\rm P}=294.88{^{\circ}}$       
        \item 3:\\ $e_{\rm A}=0.4009$, $e_{\rm P}=0.7481$, $a_{\rm A}=0.6293$, $\varpi_{\rm A}=M_{\rm A}=0{^{\circ}}$, $\varpi_{\rm P}=324.51{^{\circ}}$, $M_{\rm P}=302.12{^{\circ}}$
\end{itemize}

A detailed and exhaustive exploration of the way the asymmetry of orbits affects WD pollution is not the purpose of our present study. By relying on the linear stability of the above asymmetric periodic orbits (all of them are unstable), we directly perform the $N$-body simulations in Sect. \ref{res}, and provide a neat example of instability events that are linked with them.

\subsection{3D-RTBPs}\label{3dmaps}
Although in the literature many $N$-body investigations of post-main-sequence exoplanetary systems have so far assumed coplanarity or near-coplanarity, it is likely given from our solar system knowledge that none of the asteroids is perfectly coplanar with the planet. In order to address the physical issue of coplanarity, we therefore studied the neighborhood of a spatial periodic orbit in the 3D-CRTBP.

\begin{figure}
\centering
\resizebox{0.4\hsize}{!}{\includegraphics{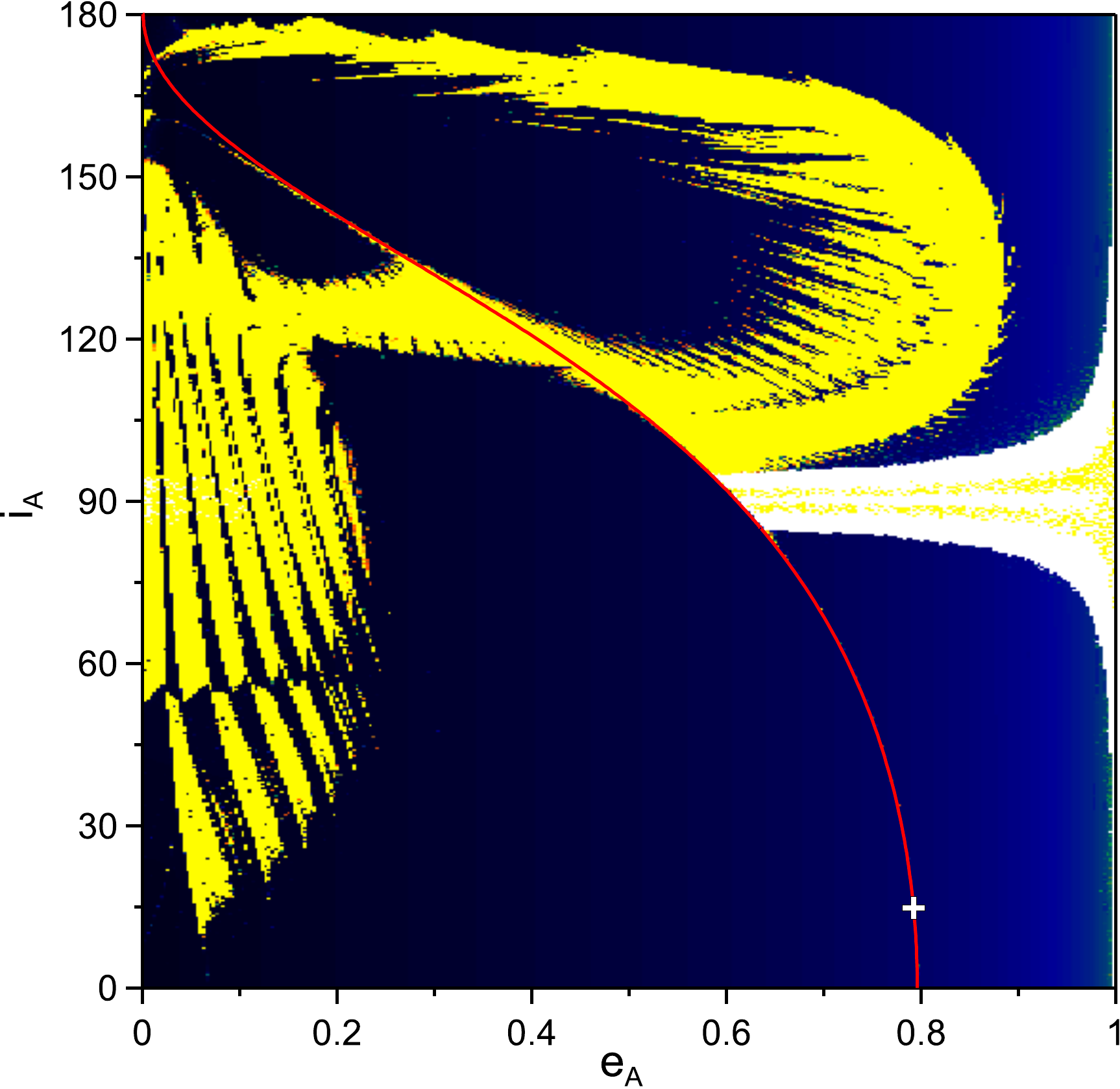}}
\caption{Family of unstable periodic orbits in 2:1 MMR of the 3D-CRTBP when $m_{\rm P}=m_{\rm J}$ is overplotted on the DS map of the plane $(e_{\rm A},i_{\rm A})$. Presentation as in Fig. \ref{21e_maps}.}
\label{21_3d_crtbp}
\end{figure}

More specifically, in Fig. \ref{21_3d_crtbp} we selected the family of unstable symmetric periodic orbits in the 3D-CRTBP ($e_{\rm P}=0$) shown by \citet{spa}. For reasons of completeness, it is presented for both prograde ($i_{\rm A}<90{^{\circ}}$) and retrograde ($i_{\rm A}>90{^{\circ}}$) orbits. The chosen periodic orbit (white cross) for the computation of the DS map on the $(e_{\rm A},i_{\rm A})$ plane has orbital elements:
\begin{itemize}
        \item $e_{\rm A}=0.7922$, $a_{\rm A}=0.6295$, $i_{\rm A}=15.03{^{\circ}}$, $i_{\rm P}=0.0{^{\circ}}$, $\varpi_{\rm A}=M_{\rm A}=\varpi_{\rm P}=M_{\rm P}=0{^{\circ}}$
        \end{itemize}

With regard to the 3D-ERBTP, we only focused on the effect of the vertical stability of the four planar symmetric periodic orbits (white crosses) selected in Fig. \ref{21e_maps}. We did not study the spatial family of symmetric periodic orbits emanating from the magenta dot (v.c.o. linking the 2D-ERTBP with the 3D-ERTBP, see \citet{spa} for its depiction) in Fig. \ref{21e_maps} because it consists of stable spatial periodic orbits.

\section{$N$-body simulations}\label{nbody}
\subsection{Integrator setup}
We used the Bulirsch-Stoer integrator from the modified version of {\sc mercury} \citep{chambers1999} and imposed an accuracy of $10^{-13}$. This provides conservation of energy and angular momentum in the range of $10^{-8}$ -- $10^{-12}$. We also configured the ejection radius to be at $3 \times 10^5$ au. 

We ran simulations with sets of 10-50 massless asteroids for different fixed combinations of orbital elements and for up to 14 Gyr, as in \cite{wd1}. Many simulations did not run for that long because of the increased parameter space and limited computational resources. Even just 10 Gyr is still older than every known polluted WD \citep{holetal2018}. The output frequency was 1 Myr.

\subsection{Choice of parameters}
In addition to the periodic orbits, which expedite the guesswork by providing the initial conditions in order to search the phase space, $N$-body integrations were our second means to help us determine the orbital evolution over long time spans.
The link between our two tools has to be carefully determined, so that the initial parameters given by the periodic orbits are correctly transformed so that the real system, where the mass of the WD is $m_{\rm WD}=0.6 m_{\odot}$, can be accurately studied and represented. In order to convert the data given by the periodic orbits into real units and thus incorporate the periodic orbits into the simulations, we therefore introduce a scaling factor for the sake of the invariance of the equations of motion for the periodic orbits. 

We performed the same scaling as was introduced in \cite{wd1} with regard to the semimajor axes used in the simulations,
\begin{equation}
a_{\rm P}^{\rm (N)} = a_{\rm P} \zeta^{1/3}\;\; {\rm and} \;\; a_{\rm A}^{\rm (N)} = a_{\rm A} \zeta^{1/3}
,\end{equation}
 where $\zeta \equiv \frac{m_{\rm WD} + m_{\rm P}}{1m_{\odot}}$ is the scaling factor and $\rm N$ represents the orbital elements used in the $N$-body simulations. The rest of the orbital elements and the time remained the same.

Furthermore, in all cases we set $a_{\rm P} = 10$ au (or scaled as $a_{\rm P}^{\rm (N)} = 8.439009789$ au). As the internal properties of the bodies are beyond the scope of our study, we adopted a fictitious WD radius equal to $R_{\rm WD}=10^6$ km $\approx 1.4378 R_{\odot}$ and a planetary radius corresponding to $R_{\rm P}=78,000$ km or 1.09 $R_{\rm J}$. The value of $R_{\rm WD}$ conforms well to a conservatively large estimate of the WD Roche radius, see \cite{veretal2017}, and during our simulations, we assumed that a collision occurs with the WD when an asteroid intersects this radius.

\subsection{Instability timescales and output visualization} 

We used both colors and shapes in order to visualize different evolution outputs of our $N$-body integrations. Namely:
\begin{itemize}
\item Green circles correspond to stablility
\item Red triangles represent the ejection of the asteroid
\item Brown squares indicate the collision between the asteroid and the planet
\item Purple diamonds signify the collision between the asteroid and the WD 
\end{itemize} 

For the stable simulations, we used a color-shading range of $[10^7, 1.4\times 10^{10}]$ yr, and for the other simulations, we used a color-shading range of $[10^3, 1.4\times 10^{10}]$ yr. The darker the shading, the longer this simulation ran.

We note that in the $N$-body integrations stability only reflects the absence of ejection of the asteroid, or collision with the planet or the WD. Even though an evolution of the orbital elements might be irregular (within the neighborhood of unstable periodic orbits), their initial values might not change significantly over time, as is the case of weakly chaotic orbits, which are considered as stable from a physical point of view. This near-invariance suggests that the output of the $N$-body integration would then be classified as stable in the adopted visual representation of the $N$-body simulation outcomes.

\section{Results of $N$-body simulations}\label{res}
In this section we illustrate the agreement between the outcome of the $N$-body simulations and the DS maps based on the periodic orbits by presenting a selection of results. Based on the DS maps shown in Sect. \ref{tools}, we chose  the initial conditions for which the DFLI traced chaoticity (DFLI>15). Then, we plot these initial conditions in the background using a uniform color for all of them and overplot the outcomes of the $N$-body simulations in order to compare and contrast the results derived from our tools. 

\subsection{2D-RTBPs}
\begin{figure}
\centering
\includegraphics[width=9cm]{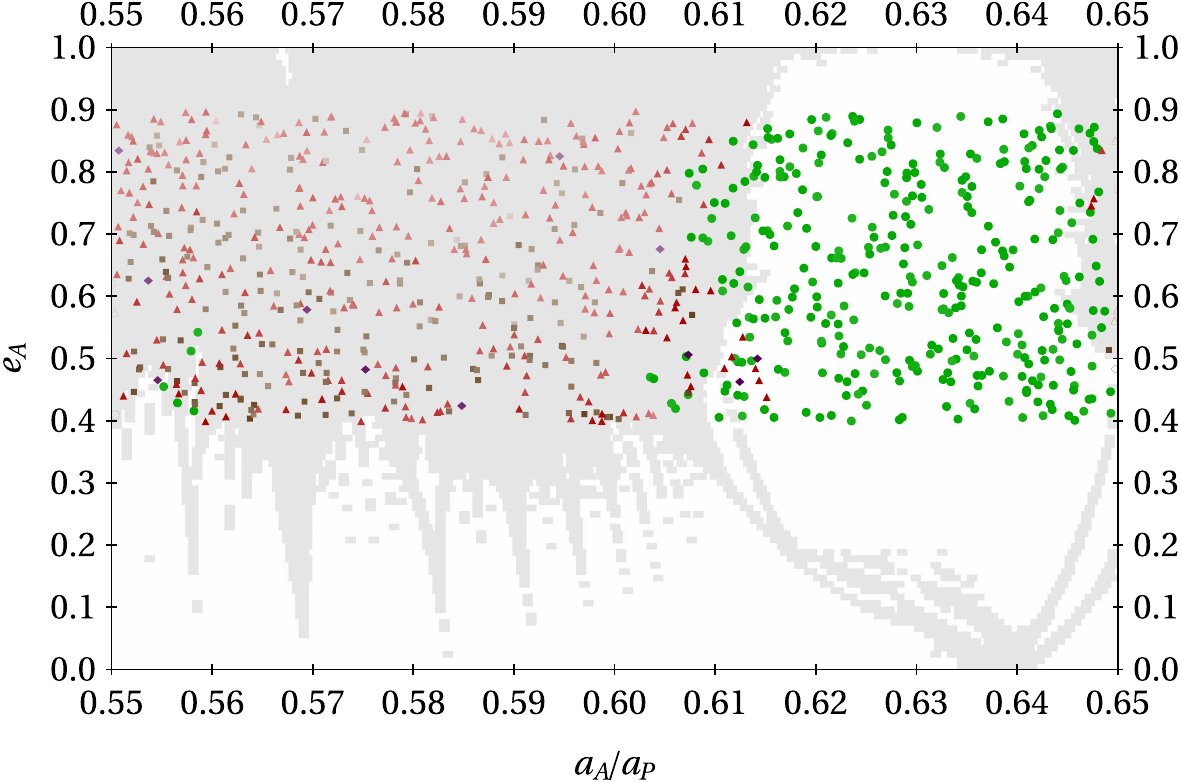}\\
\includegraphics[width=9cm]{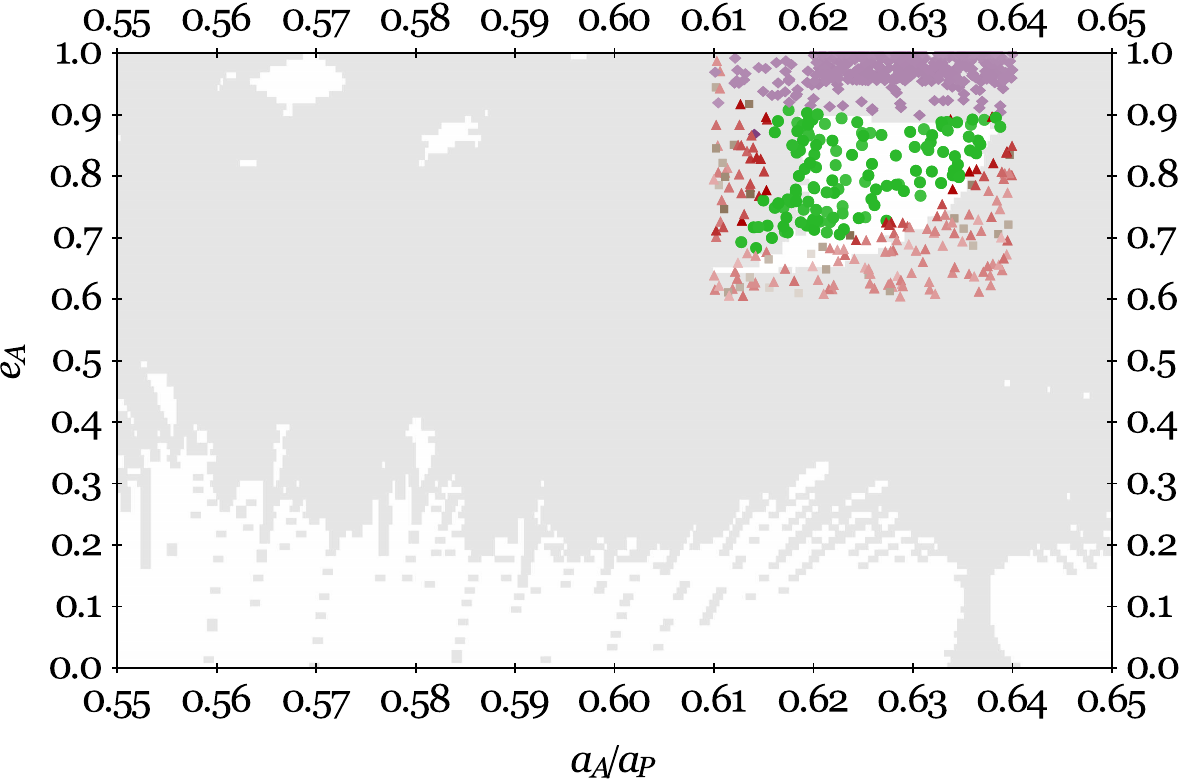}\\
\includegraphics[width=9cm]{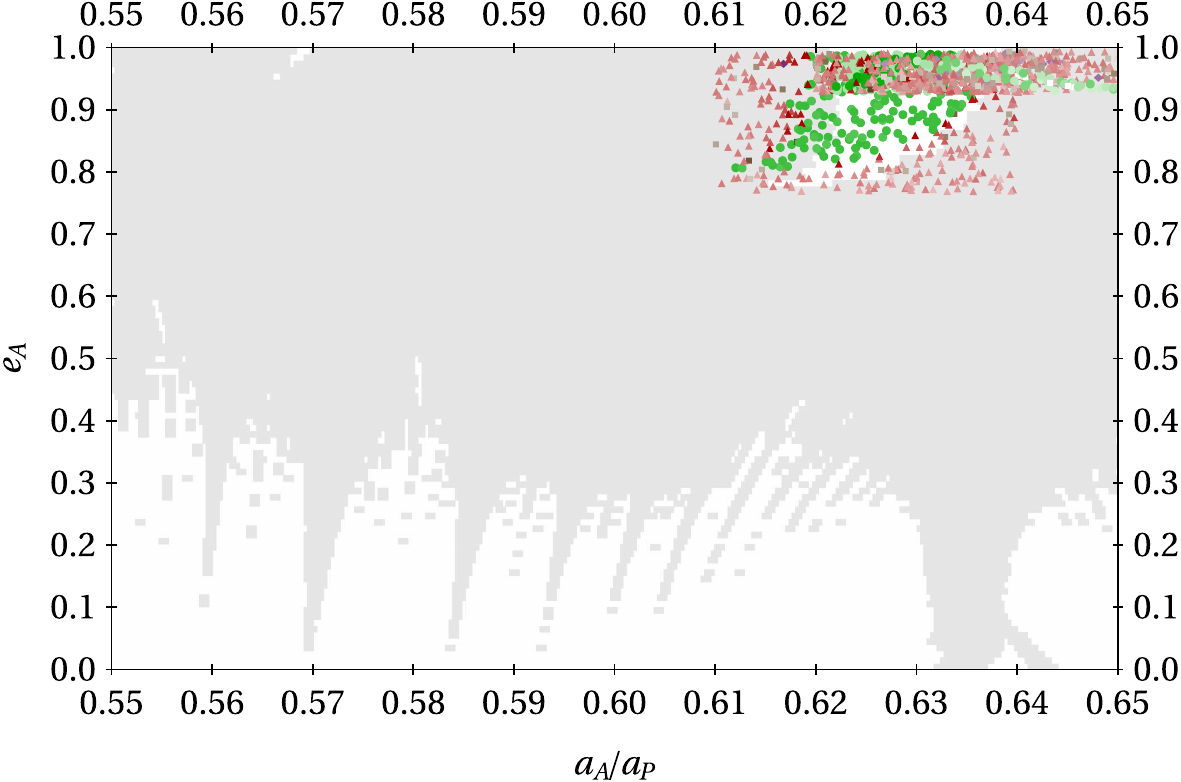}\\ \vspace{0.2cm}
\caption{Results of the $N$-body simulations on the plane $(a_{\rm A}/a_{\rm P},e_{\rm A})$ for the symmetric configuration ($\theta_1,\theta_2)$=$(0,0)$ and $\Delta\varpi$=0 (top), ($\theta_1,\theta_2)$=$(\pi,0)$ and $\Delta\varpi$=$\pi$ (middle) and ($\theta_1,\theta_2)$=$(\pi,\pi)$ and $\Delta\varpi$=0 (bottom). Green circles stand for stable outcome, red triangles for an ejected asteroid, brown squares for an asteroid that hit the planet and purple diamonds for an asteroid that collided with the WD. In the background, the initial conditions with DFLI>15 are included and uniformly depicted by pale gray.}
\label{nbody_conf_ae}
\end{figure}

We here compare the chaotic domains around unstable periodic orbits shown in Sect. \ref{2dmaps} and the $N$-body simulations for the 2D-RTBPs for the symmetric and asymmetric cases. More specifically, in Fig. \ref{nbody_conf_ae} we present the results on the plane $(a_{\rm A}/a_{\rm P},e_{\rm A})$ for the configurations ($\theta_1,\theta_2)$=$(0,0)$ (top panel), ($\theta_1,\theta_2)$=$(\pi,0)$ (middle panel), and ($\theta_1,\theta_2)$=$(0,0)$ (bottom panel). \citet{debes12} showed that the majority of tidally disrupted asteroids arise from the boundaries of the libration width of 2:1 MMR (Fig. 2 therein). \cite{picetal2017} concluded that of the 2:1, 3:1, and 4:1 MMRs, only the 4:1 MMR could excite the eccentricity of the asteroid enough for it to be disrupted by falling onto the central star. By searching for an increased concentration of purple diamonds in each of the panels of Fig. \ref{nbody_conf_ae}, we observe that the configuration ($\theta_1,\theta_2)$=$(\pi,0)$ and $\Delta\varpi$=$\pi$ (middle panel) can generate WD pollution for the orbits close to the 2:1 MMR.

\begin{figure}
\centering
\includegraphics[width=9cm]{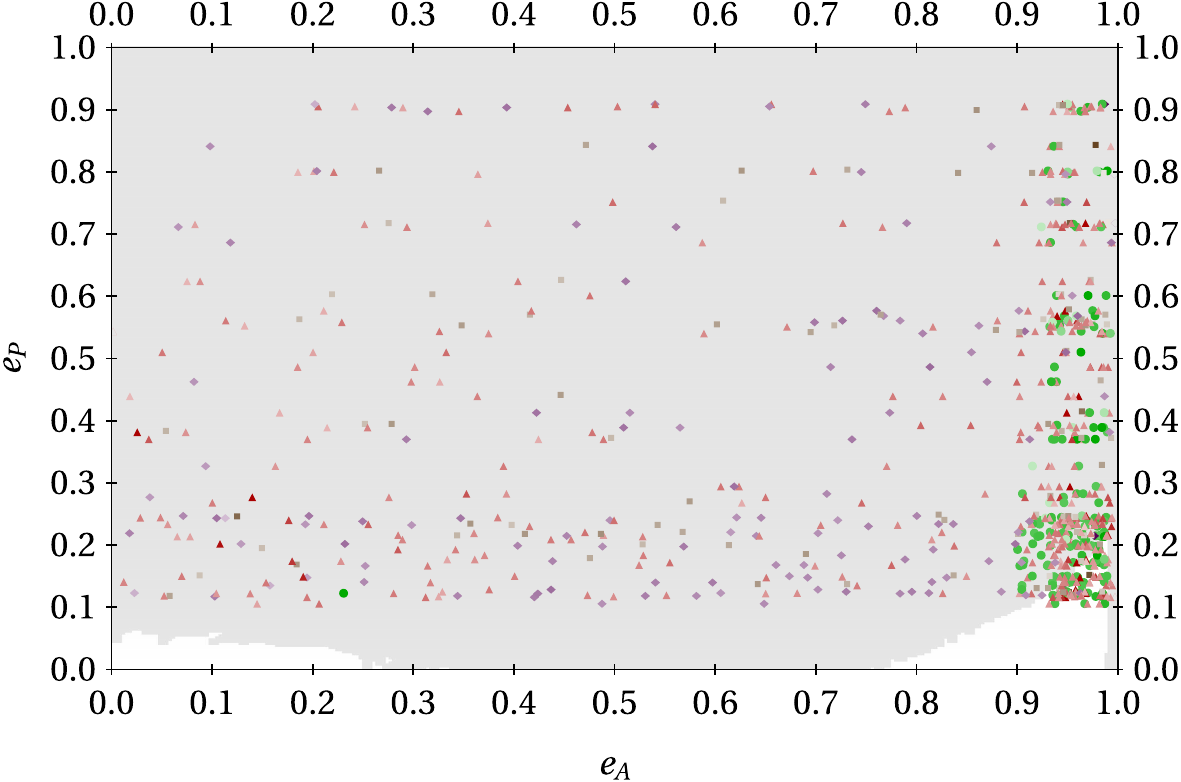}\\ \vspace{0.2cm}
\includegraphics[height=7cm]{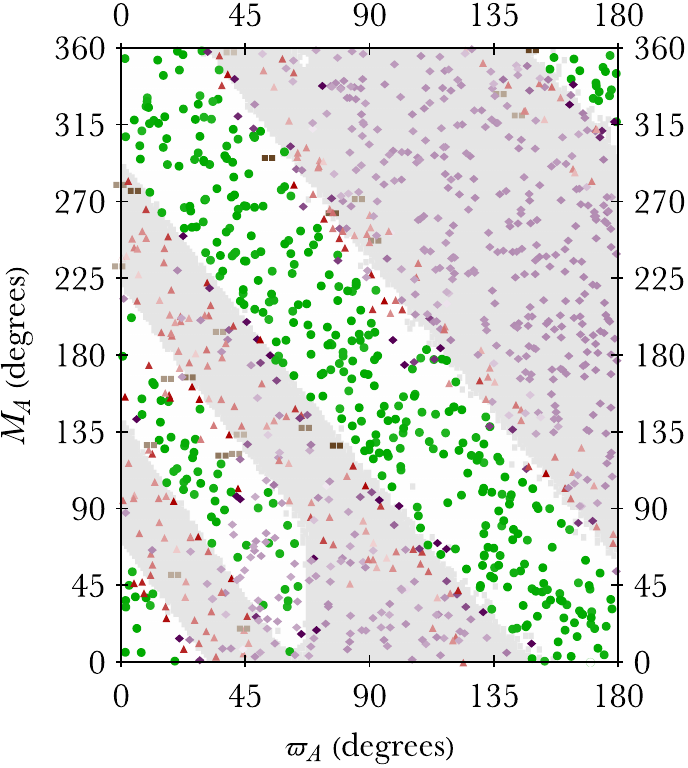}\vspace{-0.2cm}
\caption{Results of the $N$-body simulations for the symmetric configuration ($\theta_1,\theta_2)$=$(\pi,\pi)$ and $\Delta\varpi$=0 on the planes $(e_{\rm A},e_{\rm P})$ (top) and $(\varpi_{\rm A},M_{\rm A})$ (bottom).}
\label{nbody_confpp_ewMA}
\end{figure}

In Fig. \ref{nbody_confpp_ewMA} we present the results for the configuration where ($\theta_1,\theta_2)$=$(\pi,\pi)$ and $\Delta\varpi$=0 on the planes $(e_{\rm A},e_{\rm P})$ (top panel) and $(\varpi_{\rm A},M_{\rm A})$ (bottom panel). The top panel shows that the tidal disruption of the asteroid can take place for any value of $e_{\rm A}$ when $e_{\rm P}>0.1$. Additionally, we observe the remarkable role of the angles in the evolution of the asteroid, as the bottom panel is densely populated by purple diamonds when the stability region is avoided.

\begin{figure}
\centering
\includegraphics[width=9cm]{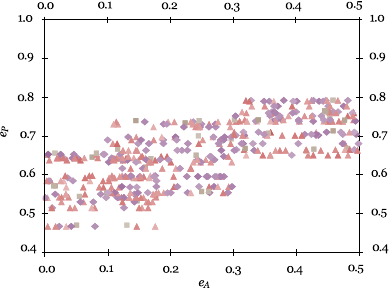}\vspace{-0.2cm}
\caption{Results of the $N$-body simulations for the three asymmetric configurations presented on the $(e_{\rm A},e_{\rm P})$-plane.}
\label{nbody_asym}
\end{figure}

Because of the strong dependence of stability on orbital angles, we wished to explore their effect by examining the neighborhood of unstable asymmetric periodic orbits, whose angles vary within the family they belong to and are not constant, like for the symmetric periodic orbits. In Fig. \ref{nbody_asym} we therefore merged the results of the $N$-body simulations for the three asymmetric configurations. In none of them was a stable evolution observed, while in the majority of the simulations, the asteroid either hit the WD or was ejected.

\subsection{3D-RTBPs}
The aim of this section is to provide clues when architectures with inclined asteroids are considered.
First we consider the 3D-CRTBP. Based on the DS map of Fig. \ref{21_3d_crtbp}, the $N$-body simulations are not encouraged because the region of instability around the unstable family is very small for prograde orbits. This means that the 3D-CRTBP, when the asteroid inclination is small, cannot trigger WD pollution, just like the 2D-CRTBP that was explored in \citet{wd1}.

\begin{figure}
\centering
\includegraphics[width=9cm]{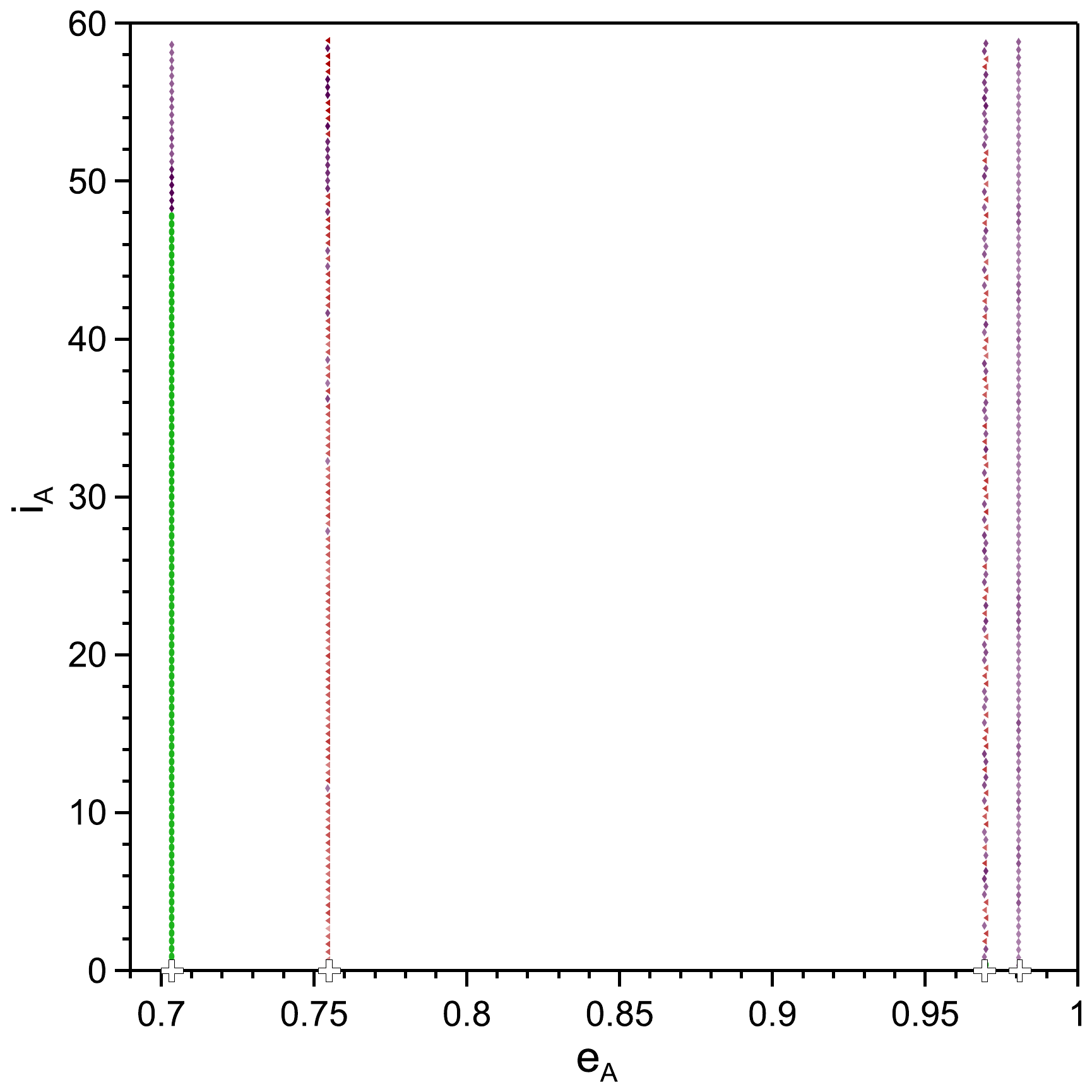}\vspace{-0.2cm}
\caption{Results of the $N$-body simulations when we considered architectures in the 3D-ERTBP for each symmetric configuration shown in Fig. \ref{21e_maps}. The white crosses represent the chosen symmetric periodic orbits with $e_{\rm P}=0.048$ from the 2D-ERTBP, which act as initial conditions for the simulations in the 3D-ERTBP, as inclination is added to the asteroids thus.}
\label{nbody_3d}
\end{figure}

With regard to the 3D-ERTBP, we described above that the neighborhood of spatial periodic orbits was not studied because in the elliptic problem, all of them were found to be stable in the 2:1 MMR. From a dynamical point of view, architectures in this problem can therefore only be studied when the vertical stability of the periodic orbits in the 2D-ERTBP is taken into account. In Fig. \ref{nbody_3d} we used as initial conditions the four planar symmetric periodic orbits (defined in Sect. \ref{2dmaps} and delineated by white crosses) that belong to each of the four symmetric configurations (shown in Fig. \ref{21e_maps}) and added inclination (with the longitudes of ascending node $\Omega_{\rm A}=\Omega_{\rm P}=0$), so that we study the outcome of the simulations in the 3D-ERTBP. We recall that all of these periodic orbits are vertically unstable (dashed lines in Fig. \ref{21e_maps}). Figure \ref{nbody_3d} shows that when the periodic orbit was horizontally unstable (red lines in Fig. \ref{21e_maps}), instability events were numerous, even for very low inclination values. We may say that WD pollution can be obtained through the configurations ($\theta_1,\theta_2)$=$(\pi,0)$ and ($\theta_1,\theta_2)$=$(\pi,\pi)$, while in the configuration ($\theta_1,\theta_2)$=$(0,0)$, an inclination $i_{\rm A}>48{^{\circ}}$ should be achieved first.

\section{Summary}\label{con}

The old age of white dwarf planetary systems introduces modeling challenges, particularly for long-term numerical simulations. Analytic and semianalytic models can provide time-saving alternatives while simultaneously revealing fundamental insights about structures such as the three-body problem.

Unstable periodic orbits with high eccentricity are here the analytic channel through which we studied the prospect of an eccentric planet that perturbs an asteroid into a white dwarf, thereby polluting it with metals. These periodic orbits enabled us to compute dynamical (in)stability maps for white dwarf planetary systems consisting of one giant planet with the same eccentricity and mass as Jupiter\footnote{Future applications may include studies with eccentricity and mass values of confirmed Jovian exoplanets.}, plus an internal asteroid (Figs. 1-6). Supplemented with a suite of computationally expensive $N$-body simulations, the maps provide a useful diagnostic for identifying initial conditions that can generate white dwarf metal pollution in 2:1 MMR. 

In order to link polluted white dwarfs with architectures that arise from planet formation, the initial conditions corresponding to the purple diamonds in Figs. 7-10 could represent final-state targets for simulations of planetary systems that traverse the main-sequence and giant branch phases.  In a comprehensive novel work, \cite{haretal2018} made the important chemical association between protoplanetary-disk formation locations and condensation temperatures with particular white dwarf pollutants. A next step could be to match the chemical connection with a planet-asteroid dynamical association, and the dynamical maps presented here may facilitate this task.

A narrower but still important application of our dynamical maps would be to explore how the accretion rate onto white dwarf systems would change if other massive bodies were present in the system in addition to the one planet considered here. For example, major planets that lie close to the Roche radius of the white dwarf would scatter incoming asteroids, either enhancing or lowering the accretion rate compared to the one-planet case. The two minor planets already found at or within the Roche radius \citep{vanetal2015,manetal2019} have prompted additional searches for major planets, and demonstrates that planetary orbits that lie entirely within one solar radius can be achieved.

Finally, the results provided in this work have applications in the broader context of celestial mechanics, and in particular for any architecture that can be efficiently modeled by the 2D-ERTBP, 3D-CRTBP, and 3D-ERTBP. Examples range from Earth-based architectures (planet-spacecraft-satellite, where the spacecraft is massless), to solar system-based architectures (Sun-Earth-Jupiter, where Earth is massless) to extrasolar system-based architectures (star-star-circumprimary planet, where the planet is considered massless).\\

\begin{flushleft}
{\bf{Acknowledgements}}
\end{flushleft}
We thank the referee for their helpful comments that have improved the manuscript. Computational resources have been provided by the Consortium des \'Equipements de Calcul Intensif, funded by the Fonds de la Recherche Scientifique de Belgique (F.R.S.-FNRS) under Grant No. 2.5020.11 and by the Walloon Region. DV gratefully acknowledges the support of the STFC via an Ernest Rutherford Fellowship (grant ST/P003850/1).

\end{document}